\documentclass[journal=jpccck,manuscript=article]{achemso}
\usepackage{multirow}
\usepackage{chemformula} 
\usepackage[T1]{fontenc} 
\usepackage{hyperref}
\usepackage{comment}
\usepackage{siunitx} 

\author{Emanuel J. A. dos Santos}
\affiliation[IFUnB]
{University of Bras\'ilia, Institute of Physics, 70910-900, Bras\'ilia, Federal District, Brazil.}
\alsoaffiliation[LCCMat]
{Computational Materials Laboratory, LCCMat, Institute of Physics, University of Bras\'ilia, 70910-900, Bras\'ilia, Federal District, Brazil.}

\author{Rodrigo A. F. Alves}
\affiliation[IFUnB]
{University of Bras\'ilia, Institute of Physics, 70910-900, Bras\'ilia, Federal District, Brazil.}
\alsoaffiliation[LCCMat]
{Computational Materials Laboratory, LCCMat, Institute of Physics, University of Bras\'ilia, 70910-900, Bras\'ilia, Federal District, Brazil.}

\author{Alexandre C. Dias}
\affiliation[CIF]{Institute of Physics and International Center of Physics, University of Bras{\'{i}}lia, 70919-970, Bras{\'{i}}lia, Federal District, Brazil.}

\author{\\Marcelo L. Pereira, Jr}
\affiliation[ENE]
{University of Bras\'{i}lia, College of Technology, Department of Electrical Engineering, 70910-900, Bras\'{i}lia, Federal District, Brazil.}

\author{Douglas S. Galvão}
\affiliation[Unicamp]
{Department of Applied Physics and Center for Computational Engineering and Sciences, State University of Campinas, 13083-859, Campinas, São Paulo, Brazil.}

\author{Luiz A. Ribeiro, Jr}
\affiliation[IFUnB]
{University of Bras\'ilia, Institute of Physics, 70910-900, Bras\'ilia, Federal District, Brazil.}
\alsoaffiliation[LCCMat]
{Computational Materials Laboratory, LCCMat, Institute of Physics, University of Bras\'ilia, 70910-900, Bras\'ilia, Federal District, Brazil.}
\email{ribeirojr@unb.br}

\title[Silverene and Copperene]
  {Exploring Novel 2D Analogues of Goldene: Electronic, Mechanical, and Optical Properties of Silverene and Copperene}

\abbreviations{IR,NMR,UV}
\keywords{American Chemical Society, \LaTeX}

\begin{document}

\begin{tocentry}
\begin{center}
\includegraphics[width=0.7\linewidth]{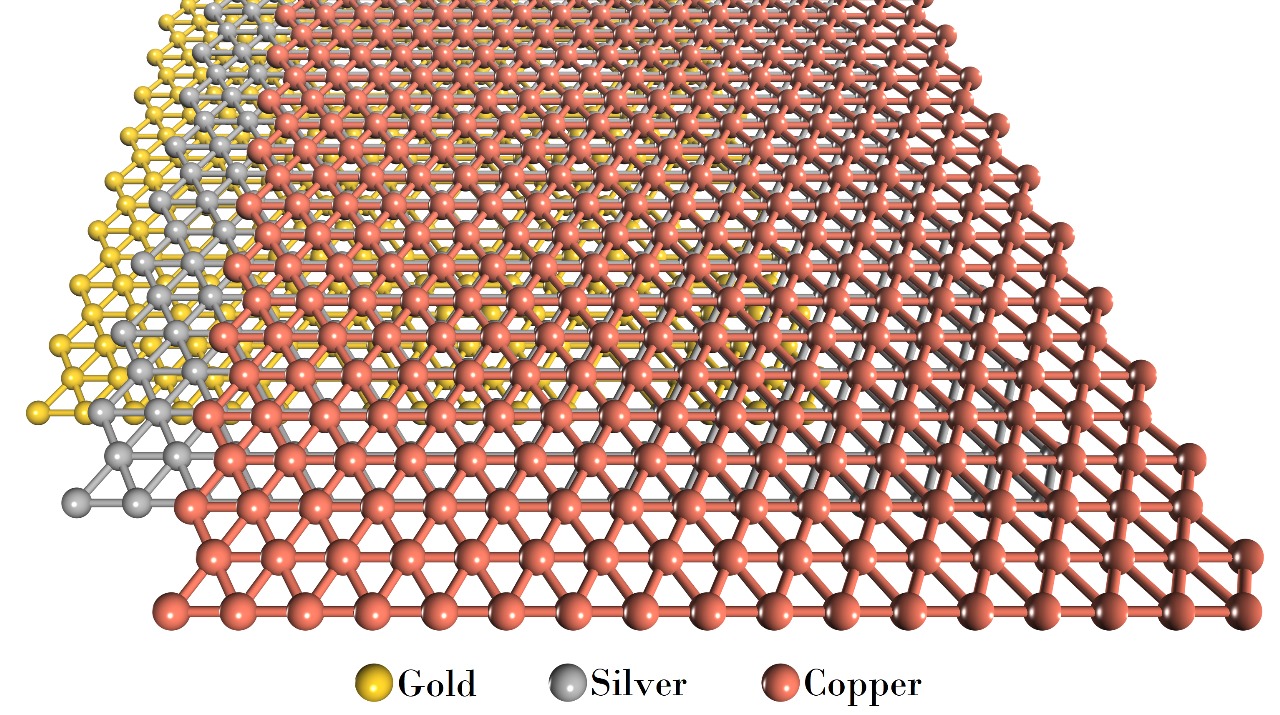}    
\end{center}
Proposed 2D copper and silver monolayers from goldene.
\end{tocentry}

\begin{abstract}
\noindent Two-dimensional (2D) materials have garnered significant attention due to their unique properties and broad application potential. Building on the success of goldene, a monolayer lattice of gold atoms, we explore its proposed silver and copper analogs, silverene and copperene, using density functional theory calculations. Our findings reveal that silverene and copperene are energetically stable, with formation energies of -2.3 eV/atom and -3.1 eV/atom, closely matching goldene's -2.9 eV/atom. Phonon dispersion and ab initio molecular dynamics simulations confirm their structural and dynamical stability at room temperature, showing no bond breaking or structural reconfiguration. Mechanical analyses indicate isotropy, with Young's moduli of 73 N/m, 44 N/m, and 59 N/m for goldene, silverene, and copperene, respectively, alongside Poisson's ratios of 0.46, 0.42, and 0.41. These results suggest comparable rigidity and deformation characteristics. Electronic band structure analysis highlights their metallic nature, with variations in the band profiles at negative energy levels. Despite their metallic character, these materials exhibit optical properties akin to semiconductors, pointing to potential applications in optoelectronics.
\end{abstract}

\section{Introduction}

Two-dimensional (2D) materials have revolutionized nanoscience, offering unparalleled properties and applications across electronics \cite{lemme20222d}, photonics \cite{xia2014two}, catalysis \cite{deng2016catalysis}, and energy storage \cite{pomerantseva2017two}. Graphene \cite{geim2009graphene}, for instance, opened avenues for exploring other 2D systems, such as borophene \cite{ranjan2020borophene}, phosphorene \cite{carvalho2016phosphorene}, and silicene \cite{kara2012review}, each exhibiting unique mechanical, electronic, and optical characteristics. Among metallic 2D materials, goldene, a single-atom-thick monolayer of gold, has emerged as a significant breakthrough, demonstrating good structural stability with roughly 9\% lattice contraction compared to bulk gold \cite{kashiwaya2024synthesis,sharma2022synthesis}. Goldene was initially synthesized using various approaches, including thermal dewetting of thin gold films and exfoliation from Au-intercalated MAX (Ti$_3$SiC$_2$) phases \cite{kashiwaya2024synthesis}. These techniques established the material’s energetic, structural, and dynamic stability. Computational studies revealed goldene's robust metallic properties, unique optical band gap, and flexibility \cite{mortazavi2024goldene,zhao2024electrical,nguyen2025goldene,abidi2024gentle,junior2025does,berdiyorov2025robust,Abidi_1_2025}. 

A recent study has further highlighted the remarkable conductivity of goldene, emphasizing its potential as a high-performance 2D conductor \cite{ramachandran2024gold}. It was demonstrated through first-principles calculations that goldene exhibits intrinsic electrical conductivity on par with lightly doped graphene, significantly surpassing other 2D metallic materials, including MXenes and MBenes \cite{mortazavi2024goldene}. This outstanding conductivity is attributed to goldene's large Fermi velocity and simple electronic structure with a prominent single energy band crossing the Fermi level. Furthermore, goldene displays high thermal conductivity, adhering closely to the Wiedemann-Franz law, which underscores its suitability for applications in next-generation nanoelectronics \cite{zhao2024electrical}. 

Beyond its intrinsic properties as a 2D material, goldene exhibits versatility by serving as a precursor to one-dimensional (1D) systems such as goldene nanotubes \cite{ono2024breakdown}. Recent studies have demonstrated that rolling goldene into nanotube geometries produces goldene nanotubes (GNTs) with distinctive electronic and mechanical properties. GNTs retain the metallic nature of goldene while exhibiting unique electronic features, such as nearly flat bands near the Fermi level for specific chirality indices. These insights into the electronic and structural properties of goldene strengthen the case for exploring analogous materials, such as silverene and copperene, which could exhibit comparable or superior properties.

In this study, we propose and computationally investigate the 2D analogs of goldene, named silverene and copperene. Using density functional theory (DFT) calculations, we explore their structural, mechanical, electronic, and optical properties. We establish their energetic and dynamic stability through formation energy calculations, phonon dispersions, and \textit{ab initio} molecular dynamics simulations. Additionally, we evaluate their isotropic mechanical characteristics, including Young's modulus and Poisson's ratio, and analyze their electronic band structures to highlight similarities and differences with goldene. Despite their metallic nature, goldene, silverene and copperene exhibits optical properties akin to semiconductors, indicating a potential for optoelectronic applications.

\section{Methodology}

This study used density functional theory (DFT) calculations to investigate the structural, electronic, and optical properties of copperene, goldene, and silverene monolayers. These calculations were performed using the Vienna Ab initio Simulation Package (VASP) \cite{kresse1993ab,kresse1996efficient}, employing the generalized gradient approximation (GGA) with the Perdew-–Burke-–Ernzerhof (PBE) exchange-correlation functional \cite{perdew1996generalized}. The Kohn-Sham equations were solved using the plane-wave basis set in combination with the projector augmented wave (PAW) method \cite{Blochl_17953_1994,Kresse_1758_1999}, which ensures accurate treatment of core and valence electron interactions. 

Geometric optimizations were performed by minimizing the stress tensor and atomic forces, utilizing a plane-wave cutoff energy of \SI{500}{\electronvolt}, for the density of states, electronic band structure, phonon dispersion and  \textit{ab initio} molecular dynamics (AIMD) we used a cutoff energy of \SI{300}{\electronvolt}. The self-consistency cycle adopted a total energy convergence criterion of \SI{E-6}{\electronvolt}, while equilibrium structures were confirmed when the residual interatomic forces were below \SI{0.01}{\electronvolt\per\angstrom}. To eliminate spurious interactions between the monolayers and their periodic images along the non-periodic z-direction, a vacuum spacing of \SI{15}{\angstrom} was introduced in the unit cells. For the structural optimization and electronic properties we used a \textbf{k}-mesh of $28\times16\times1$, generated using the Monkhorst--Pack method \cite{monkhorst1976special}.

The dynamical stability was investigated through phonon dispersion, combining VASP and Phonopy package \cite{Togo_1_2015}, using density functional perturbation theory (DFPT) \cite{Gajdos_045112_2006}, these calculations were formed using a $7\times7\times1$ supercell, with a $2\times2\times1$ \textbf{k}-points mesh. The thermodynamic properties was obtained from phonon dispersions with a a $48\times48\times1$ \textbf{k}-points mesh. The thermodynamic stability was investigated through AIMD, using NVT ensamble, with Langevin thermostat \cite{Hoover_1818_1982,Evans_3297_1983}, at \SI{300}{\kelvin} with a timestep of \SI{1}{\fs} for a simulation time of \SI{10}{\ps}. These simulations were conducted with $6\times6\times1$ supercell and considering only $\Gamma$ \textbf{k}-point.

The linear optical response were investigated at independent particle approximation (IPA) and Bethe--Salpeter equation (BSE) \cite{Salpeter_1232_1951} levels of theory via WanTiBEXOS code \cite{dias2023wantibexos}. The single-particle electron and hole energy levels were directly obtained by a maximally localized Wannier function tight binding (MLWF-TB) Hamiltonian, directly obtained from DFT calculations, using Heyd--Scuseria--Ernzerhof (HSE06) hybrid exchange-correlation function \cite{heyd_1187_2004}, for a more accurate description of the interband separations, through Wannier90 package \cite{mostofi2008wannier90}. These calculations were done using a $44\times25\times1$ \textbf{k}-points mesh, considering the following number of conduction ($n_c$) and valence ($n_v$) bands: $n_c = 3$ and $n_v =6$ for goldene,  $n_c = 4$ and $n_v =6$ for silverene and $n_c = 2$ and $n_v =3$ for copperene which are sufficient to describe the absorption spectrum in the solar emission range (i.e \SIrange{0}{4}{\electronvolt}) and smearing of \SI{0.05}{\electronvolt} was applied in the dielectric function calculations at both IPA and BSE levels. The BSE calculations employed a 2D truncated Coulomb potential (V2DT)\cite{Rozzi_205119_2006} to modeling the electron-hole Coulomb potential. The optical band gap was defined as the lowest direct transition and an oscillator force higher equals than \SI{0.1}{\square\angstrom}.

\section{Results}

\begin{figure}[!htb]
    \centering
    \includegraphics[width=0.8\linewidth]{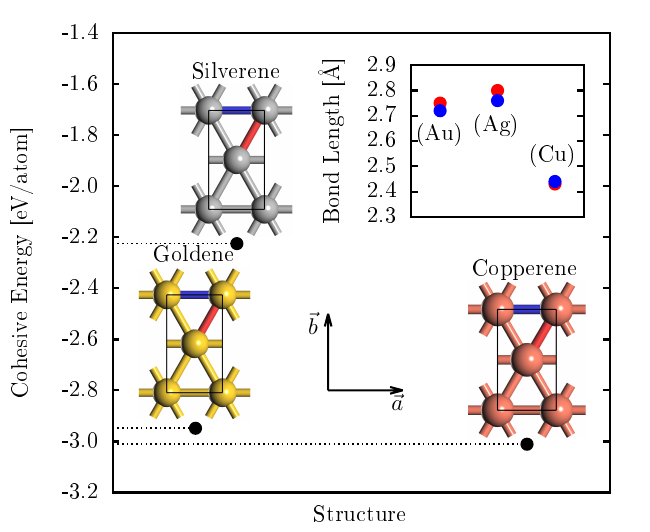}
    \caption{Optimized copperene, goldene, and silverene crystal structures, along with their cohesion energies and bond lengths. The main panel presents the cohesion energy for each material, while the inset panel compares the bond lengths (red and blue bonds) across the three monolayers.}
    \label{fig:cohesion}
\end{figure}

Goldene, silverene and copperene unit cell, together with their cohesive energies and bond lengths are shown in Fig.~\ref{fig:cohesion}. The schematic representations emphasize the structural similarities among these three materials, while red and blue bonds indicates variations in bond lengths. Additionally, the inset panel provides a detailed comparative analysis of the bond lengths, enabling a direct evaluation of their differences across the three structures.

The cohesion energy ($E_\text{cohe}$) of the investigated 2D materials was defined as:
\begin{equation}
E_\text{cohe} = \frac12\left(E_\text{Monolayer} - 2\cdot E_\text{Atom}\right),
\end{equation}
where $E_\text{Monolayer}$ is the monolayer total energy, the factor of two accounts for the number of atoms in the unit cell, and $E_\text{Atom}$ represents the energy of an isolated gas-phase atom (Au, Ag, or Cu). A negative $E_\text{cohe}$ indicates that the 2D configuration is energetically favorable, with more negative values corresponding to greater stability. 


Among the investigated 2D materials, copperene exhibits the lowest cohesion energy ($E_\text{cohe}=-3.1$ eV/atom), followed by goldene ($E_\text{cohe}=-2.9$ eV/atom) and silverene ($E_\text{cohe}=-2.3$ eV/atom), as shown in Fig.~\ref{fig:cohesion}. This trend highlights the role of interatomic interactions and bonding environments in determining stability. The enhanced stability of copperene arises from its stronger interatomic forces, whereas goldene retains relatively high stability due to persistent relativistic effects. Silverene, with its weaker bonding, is the least stable among the three. The consistently negative $E_\text{cohe}$ values confirm that all monolayers are energetically more favorable than their gas-phase atomic counterparts, emphasizing the influence of electronic effects and reduced dimensionality on stability.

The inset panel in Fig.~\ref{fig:cohesion} illustrates the bond length variations for copperene, goldene, and silverene, highlighting the differences between the blue and red bonds. These disparities stem from the constituent elements' distinct electronic environments and atomic radii, which modulate the directional bonding within the rectangular lattice. The observed bond length anisotropy reflects the interplay between covalent and metallic bonding characteristics in these 2D materials, emphasizing the structural deviations induced by element-specific interactions.

In copperene, the blue and red bonds exhibit nearly identical lengths (\SI{2.44}{\angstrom} and \SI{2.43}{\angstrom}), indicating a highly uniform bonding environment. This near-isotropic lattice structure arises from copper's strong interatomic interactions and relatively small atomic radius, contributing to its high cohesion energy and structural stability. Goldene, in contrast, presents a more pronounced bond length disparity (\SI{2.72}{\angstrom} and \SI{2.75}{\angstrom}) compared to copperene. This difference is primarily attributed to relativistic effects in gold \cite{pyykko1979relativity,schwerdtfeger2002relativistic}, which stabilize the $6s$ electrons and modify the bonding environment \cite{pyykko1979relativity}. These effects enhance the anisotropy in bond lengths, leading to more significant variations within the rectangular lattice. Silverene exhibits the most significant bond length disparity among the three materials (\SI{2.76}{\angstrom} and \SI{2.80}{\angstrom}). This increased anisotropy stems from silver's weaker interatomic bonding and larger atomic radius, resulting in a less rigid lattice than goldene and copperene. 
The bond lengths in each monolayer directly influence the lattice constants $(a_{0},b_{0})$, which are found to be (\SI{2.72}{\angstrom},\SI{4.77}{\angstrom}) for goldene, (\SI{2.75}{\angstrom}, \SI{4.87}{\angstrom}) for silverene, and (\SI{2.45}{\angstrom}, \SI{4.20}{\angstrom}) for copperene. These variations further reinforce the correlation between atomic bonding characteristics and structural anisotropy in these 2D materials.

\begin{figure}[!htb]
    \centering
    \includegraphics[width=\linewidth]{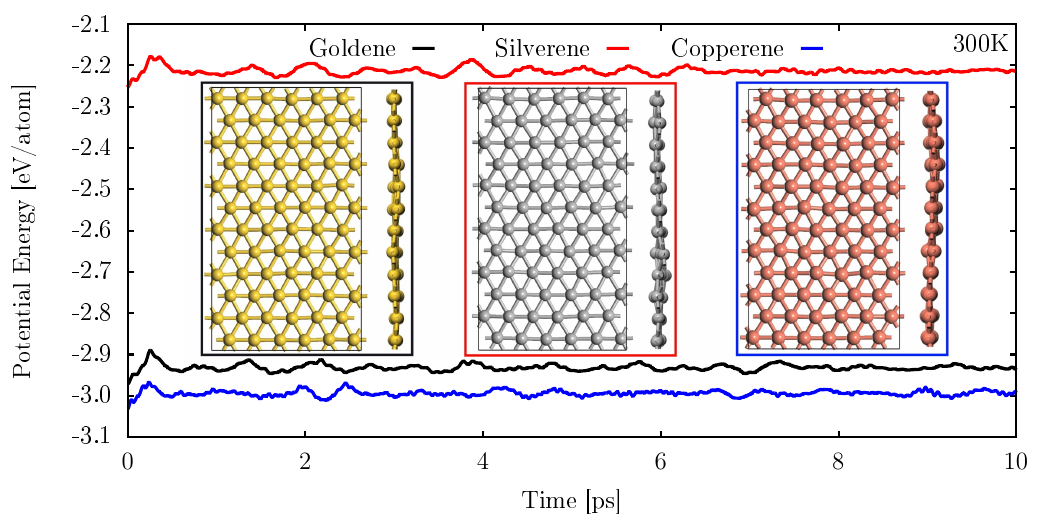}
    \caption{AIMD simulations for goldene, silverene, and copperene at 300 K. The main panel shows the time evolution of the potential energy per atom, plotted in black, red, and blue for goldene, silverene, and copperene, respectively. The inset panels display the top and side views of the final snapshots of the structures after 10 ps of simulation.}
    \label{fig:aimd}
\end{figure}

The thermodynamic stability of the investigated monolayers are shown through AIMD simulations, performed at \SI{300}{\kelvin}, in Fig.~\ref{fig:aimd}. The main panel shows the time evolution of the potential energy per atom for the three monolayers, plotted in blue, black, and red for copperene, goldene, and silverene, respectively. The inset panels display the top and side views of the final snapshots of the structures after \SI{10}{\ps} of simulation.

These AIMD simulations reveals that all monolayers remains stable at \SI{300}{\kelvin}, with no evidence of bond breaking or reconfiguration throughout the simulation. This stability underscores the robustness of these 2D materials under environmental conditions, making them promising candidates for several applications at room temperature operation. The potential energy per atom follows the order $\text{copperene} < \text{goldene} < \text{silverene}$, reflecting the relative stability of the materials. This order aligns with earlier cohesion energy calculations, where copperene exhibited the lowest cohesion energy, followed by goldene and silverene. The strong interatomic interactions and uniform bonding in copperene contributes to its lower potential energy, while the larger bond length disparity and weaker bonding in silverene result in a higher potential energy per atom. Goldene lies between these two extremes, influenced by relativistic effects that enhance its bonding. The time evolution of the potential energy shows minor fluctuations for all three materials, further confirming their dynamic stability during the AIMD simulations. The consistency of these results with the cohesion energy calculations provides strong evidence for the thermodynamic stability of copperene and silverene 2D monolayers.


\begin{figure}[!htb]
    \centering
    \includegraphics[width=\linewidth]{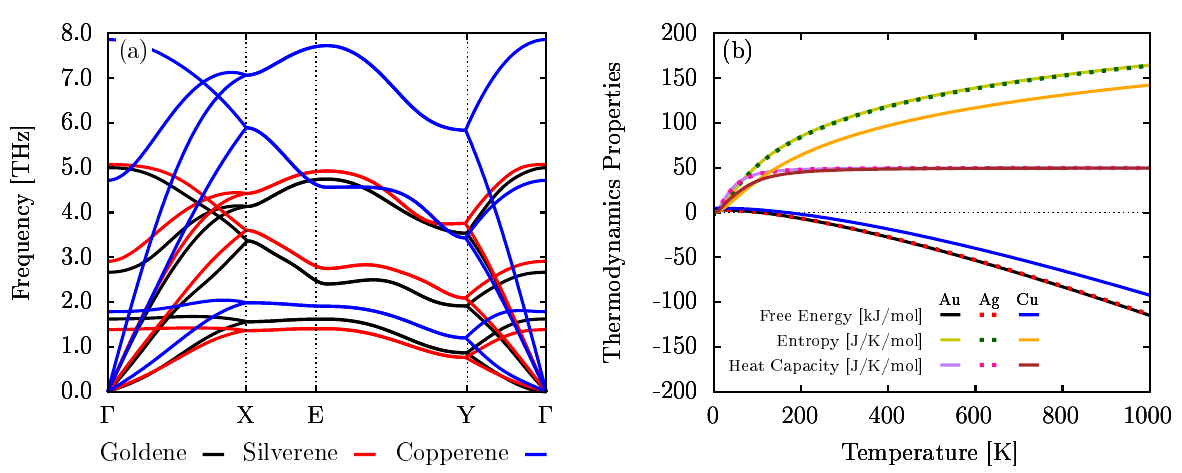}
    \caption{(a) Phonon dispersion curves for copperene, goldene, and silverene, represented in blue, black, and red, respectively. (b) Thermodynamic properties, including free energy, entropy, and heat capacity at constant volume ($C_v$), as functions of temperature.}
    \label{fig:phonon_thermo}
\end{figure}

The phonon dispersion, shown in Fig.~\ref{fig:phonon_thermo}(a), highlights key differences in the vibrational behavior of the three monolayers, the lack of imaginary frequencies also suggests the dynamical stability of these monolayers. Copperene exhibits the highest phonon frequencies, with modes reaching approximately \SI{8}{\tera\hertz}, reflecting its strong interatomic bonding and smaller atomic mass. Goldene displays intermediate phonon frequencies, influenced by relativistic effects that alter the bonding environment. Goldene's phonon dispersion is according to the other previous investigations shown in literature \cite{mortazavi2024goldene,junior2025does}. Silverene, on the other hand, has the lowest phonon frequencies, consistent with its weaker interatomic bonding. The higher frequency modes observed in copperene also suggest greater lattice stiffness than goldene and silverene. These differences directly correlate with the earlier cohesion energy and bonding strength trends, where copperene was the most stable and silverene the least stable.

Fig.~\ref{fig:phonon_thermo}(b) depicts the thermodynamic properties of the three monolayers as a function of temperature, providing insight into their thermal behavior. The free energy decreases monotonically with increasing temperature for all materials, with goldene and silverene having the most negative free energy, followed by copperene. Despite slight differences in their phonon spectra, the free energy curves for goldene and silverene are nearly identical. Being this behaviour justified by the similarities in the phonon dispersion, leading to comparable contributions to the free energy. The free energy of copperene is less negative due to its stiffer lattice, as indicated by the higher phonon frequencies, which results in reduced vibrational entropy contributions to the free energy at the same temperature.

The entropy increases with temperature for all three materials, with goldene and silverene displaying the steepest slopes. This indicates that goldene and silverene have more accessible vibrational modes at higher temperatures, consistent with their lower phonon frequencies and more flexible lattices. Copperene, on the other hand, exhibits the least entropy increase due to its stiffer lattice and fewer low-frequency modes, which limit the vibrational contributions to entropy. The heat capacity of goldene and silverene reaches its limit more quickly, around \SI{200}{\kelvin}, and remains constant for higher temperatures. This behavior results from their softer lattices and lower phonon frequency ranges, allowing them to excite their vibrational modes at lower temperatures fully. Copperene follows a similar trend but with a smaller slope, gradually reaching its limit. This difference arises from copperene's higher phonon frequencies, which delay the full excitation of its vibrational modes. The heat capacity plateau at \SI{200}{\kelvin} for all materials reflects their approach to the Dulong-Petit limit, characteristic of harmonic lattice vibrations. In the 2D systems, this saturation occurs more quickly due to the reduced dimensionality, as compared to their bulk counterparts, where the heat capacity saturates at higher temperatures due to the broader phonon density of states in the bulk \cite{mounet2005first}.

\begin{figure}[!htb]
    \centering
    \includegraphics[width=0.6\linewidth]{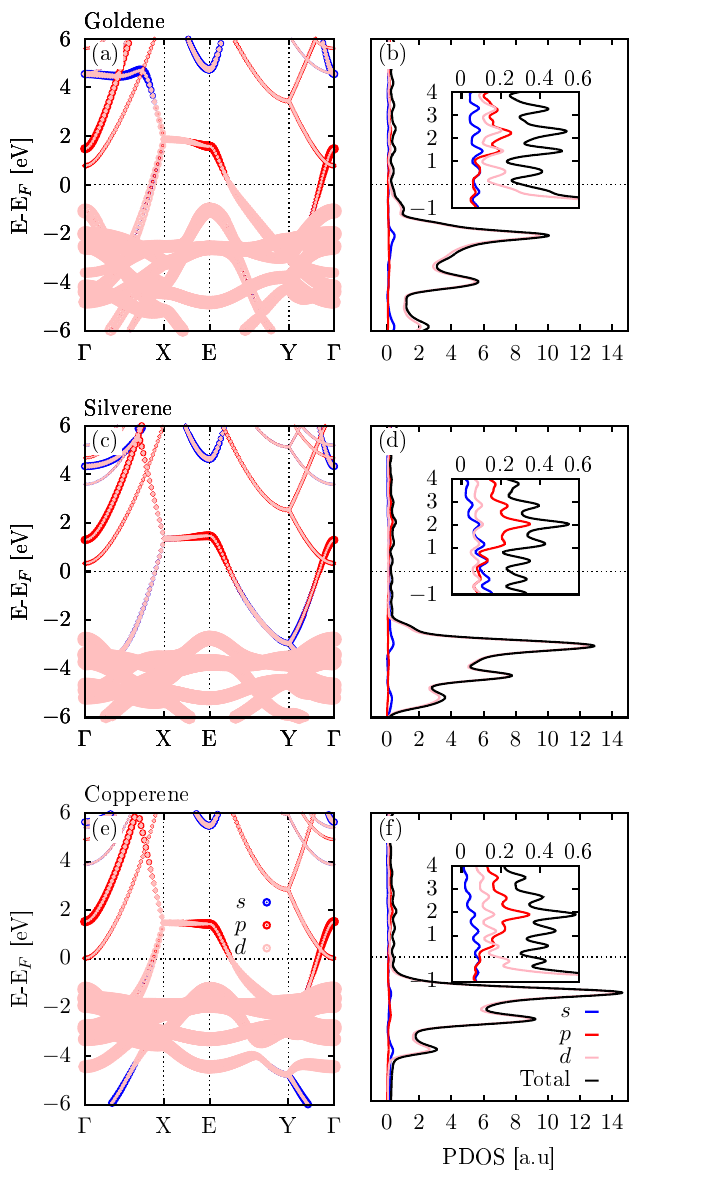}
    \caption{Electronic band structure and projected density of states (PDOS) at PBE level for (a,b) goldene, (c,d) silverene, and (e,f) copperene, shown in the fat band representation with contributions from $s$, $p$, and $d$ orbitals. The PDOS panels include an inset highlighting the energy range up to \SI{4}{\electronvolt} above the Fermi level. The Fermi level is set at \SI{0}{\electronvolt}}
    \label{fig:band_pdos}
\end{figure}

We now turn out attention to the electronic properties of the investigated materials. Fig.~\ref{fig:band_pdos} presents the electronic band structure and projected density of states (PDOS), at PBE level, for copperene, goldene, and silverene. Panels (a), (c), and (e) displays the band structure for goldene, silverene, and copperene, respectively, with fat bands highlighting the contributions of the $s$, $p$, and $d$ orbitals. Panels (b), (d), and (f) present the corresponding PDOS, with the inset panels zooming into the energy levels up to \SI{4}{\electronvolt} above the Fermi level.

The band structures and PDOS of the three materials exhibit overall similarity, with metallic characteristics evident from the band crossing the Fermi level. However, subtle differences in the band dispersions and PDOS profiles indicate distinct electronic behaviors influenced by atomic mass, bonding, and orbital contributions. A key difference is observed in the X-E path of the band structure. While silverene and copperene exhibit a flat band in this region (see Fig.~\ref{fig:band_pdos}(c,e)), goldene displays a sloping band, as illustrated in Fig.~\ref{fig:band_pdos}(a). The flat bands in silverene and copperene suggest localized electronic states, which may arise from these materials' lower atomic mass and weaker bonding interactions, when compared with goldene. This feature could enhance the density of states near these energies, potentially influencing properties like electrical conductivity and thermoelectric performance. In contrast, the slope in goldene indicates more delocalized states, contributing to enhanced electronic mobility.

One can note in Figs.~\ref{fig:band_pdos}(a,c,e) that the energy levels below Fermi level in goldene and copperene are closer to the Fermi level than in silverene. This trend can be attributed to the stronger bonding and relativistic effects in goldene and copperene, which stabilize these states closer to the Fermi energy. In silverene, the weaker bonding leads to a greater separation between the negative energy levels and the Fermi level, altering its electronic structure. Copperene exhibits a significantly higher density of states (DOS) for the occupied states region than goldene and silverene, as shown in Figs.~\ref{fig:band_pdos}(b,d,f). This behavior can be explained by the $d$ orbitals of copper contributing more prominently to the occupied states. The higher density of $d$ orbitals in this region reflects stronger localization of electrons in copperene, which could impact properties such as electron localization and bonding strength.

In Figs.~\ref{fig:band_pdos}(b,d,f), the PDOS reveals that the energy levels below the Fermi energy are predominantly constructed from $d$ orbitals. In contrast, the levels above the Fermi energy are primarily composed of $s$ and $p$ orbitals. This distinction suggests that electronic transitions between these regions may be limited by selection rules, avoiding direct transitions between $d$-dominated levels below the Fermi energy and $s/p$-dominated levels above it. These orbital characteristics can influence optical applications, particularly in determining the nature of bright and dark excitations. Bright excitations arise from allowed transitions between orbitals with overlapping symmetry and spatial distribution. On the other hand, dark excitations are associated with forbidden transitions due to symmetry/spin mismatch or weak coupling between initial and final states \cite{scholes2006excitons}. The dominance of $d$-orbitals below the Fermi level could suppress some direct transitions, favoring dark excitations, while the $s$- and $p$-orbital contributions above the Fermi level might enhance the probability of bright excitations.

\begin{figure}[!htb]
    \centering
    \includegraphics[width=\linewidth]{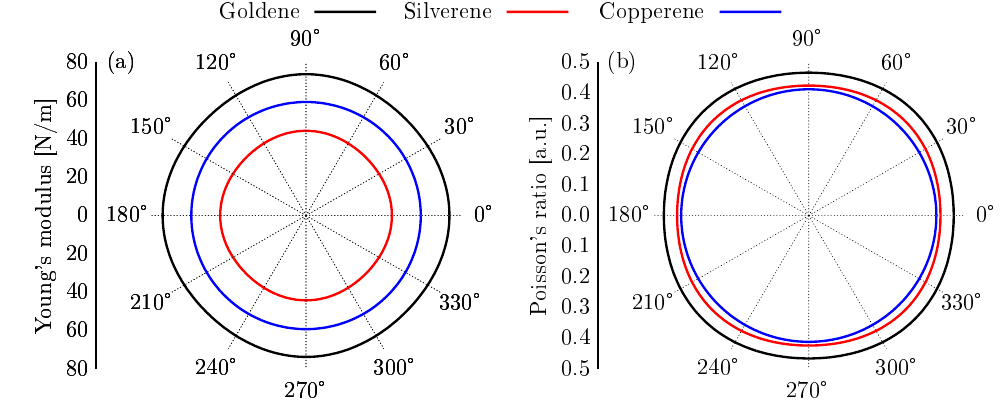}
    \caption{Polar plots of (a) Young's modulus and (b) Poisson's ratio for goldene, silverene, and copperene.}
    \label{fig:mec_prop}
\end{figure}

The mechanical properties of copperene, goldene, and silverene are also investigated. Fig.~\ref{fig:mec_prop} presents Young's modulus and Poisson's ratio in polar plots (panels (a) and (b), respectively). The near-isotropic nature of these mechanical properties is evident in the approximately circular shapes of the polar plots for both Young's modulus and Poisson's ratio. This behavior stems from the intrinsic symmetry of the rectangular crystal lattice, which results in a nearly uniform mechanical response, with only minor variations depending on the strain direction.

In Fig.~\ref{fig:mec_prop}(a), the Young's modulus values ($Y_{M}$, and their maximum values $Y_{M,{\text{max}}}$) follow the trend goldene ($Y_{M,{\text{max}}}=73$ N/m) > copperene ($Y_{M,{\text{max}}}=59$ N/m) > silverene ($Y_{M,{\text{max}}}=44$ N/m). This order is primarily dictated by the elastic constants $C_{11}$, $C_{22}$, and $C_{66}$, which are directly related to the material's resistance to deformation under stress. Goldene exhibits the highest values of $C_{11}=94.33$ N/m, $C_{12}=43.85$ N/m, $C_{22}=93.95$ N/m, and $C_{66}=24.27$ N/m, indicating its superior stiffness compared to the other materials. As mentioned above, the relativistic effects in gold enhance bonding strength, resulting in greater resistance to deformation. Copperene, with intermediate elastic constants ($C_{11}=C_{22}=71.11$ N/m, $C_{12}=29.21$ N/m, and $C_{66}=20.95$ N/m), has lower stiffness than goldene but higher than silverene. Silverene, with the lowest elastic constants ($C_{11}=53.90$ N/m, $C_{12}=22.79$ N/m, $C_{22}=53.76$ N/m, and $C_{66}=20.95$ N/m), is the most flexible due to weaker bonding and larger atomic radius.

The Poisson's ratio values follow the trend of goldene (0.47) > silverene (0.42) > copperene (0.41). This ordering is governed by the ratio between $C_{11}$ and $C_{12}$, which dictates the degree of lateral expansion when the material is compressed along a given direction. The higher Poisson's ratio in goldene is primarily attributed to its elevated $C_{12}$ value, which enhances lateral deformation. Silverene exhibits slightly more significant lateral strain than copperene, resulting in their similar Poisson's ratios, as shown in Fig.~\ref{fig:mec_prop}(b). Furthermore, all three materials exhibit Poisson's ratio values exceeding \SI{0.25}{}, indicating a predominantly ductile fracture behavior \cite{wang2019dhq}. This suggests that these 2D structures undergo significant plastic deformation under mechanical stress before failure, a characteristic associated with materials that efficiently redistribute strain energy through lateral expansion.

Finally, we present the optical absorption results.

\begin{figure}[!htb]
    \centering
    \includegraphics[width=\linewidth]{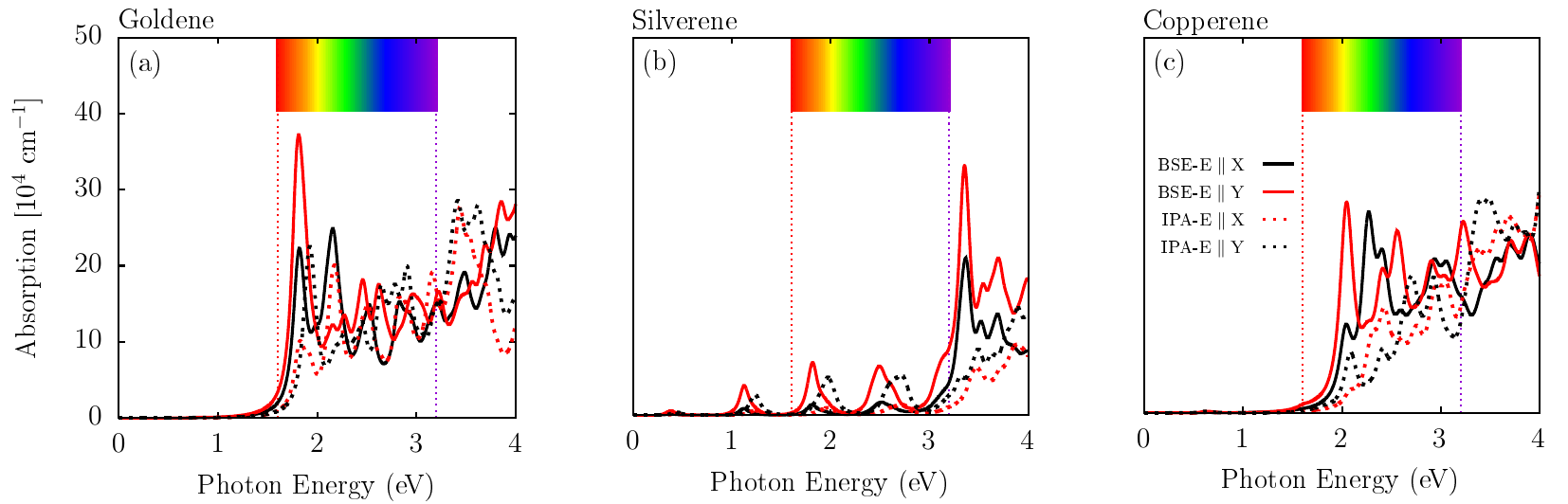}
    \caption{Optical absorption spectra for (a) goldene, (b) silverene, and (c) copperene, calculated using BSE (solid lines) and IPA (dotted lines). The spectra are shown for light polarization along the $x$-axis (black) and $y$-axis (red). The colored bar at the top maps the photon energy range to the visible spectrum.}
    \label{fig:optical_absorption}
\end{figure}

The linear optical response of the investigated monolayers are shown in Fig.~\ref{fig:optical_absorption} through the absorption coefficient using BSE and IPA levels of theory. Panels (a), (b), and (c) correspond to goldene, silverene, and copperene, respectively. The spectra are shown for light polarization along the $x$- and $y$-axis, with the solid and dotted lines representing the BSE and IPA results, respectively. The colored bar at the top maps the photon energy range to the visible spectrum, aiding in the interpretation of optical properties within the visible range.

Despite the metallic behavior in the electronic structure, these systems from the optical point of view behaves like semiconductors, as seen in Fig.~\ref{fig:optical_absorption}, being evident in their distinct absorption onsets and well-defined peaks in the visible and ultraviolet spectra for goldene and copperene and in the intra-red spectra for silverene. This can be justified by the lower population of conduction states closer to Fermi level and the different orbital symmetry of these states when compared to valence ones, as shown in PDOS Fig.~\ref{fig:band_pdos}. The combination of these two factors results in forbidden optical transitions (i.e dark excitations) or allowed transitions with a very small oscillator force (i.e lower optical transition probability rate). Due the fact that excitonic effects can results in significant red-shifts in the optical band gap of 2D materials and significant changes in the shape of absorption spectrum,\cite{Dias_3265_2021,Dias_8572_2024} we choose also to calculated these monolayer's response considering these quasi-particle effects through BSE.

The absorption spectra of goldene (Fig.~\ref{fig:optical_absorption}(a)) and copperene (Fig.~\ref{fig:optical_absorption}(c)) displays an apparent onset in the visible region. Silverene, in turn, has some small absorption activity in the infrared and visible regions, with higher activity in the ultraviolet region, as illustrated in Fig.~\ref{fig:optical_absorption}(b). Goldene exhibits the lowest absorption onset among the materials, reflecting its electronic structure with closely spaced energy levels near the Fermi level. Prominent absorption peaks are observed for goldene at approximately \SI{2.0}{\electronvolt} and \SI{3.5}{\electronvolt}. For silverene, the prominent peaks occur around \SI{2.5}{\electronvolt} and \SI{3.8}{\electronvolt}, with additional smaller features in the intermediate range. Copperene shows broader peaks near \SI{2.2}{\electronvolt} and \SI{3.5}{\electronvolt}, with an increase in overall absorption intensity compared to silverene.

The BSE results show sharper and more intense absorption peaks than IPA, particularly near the prominent peaks. This trend indicates the significant role of electron-hole interactions in enhancing light absorption, especially in goldene, where the excitonic-like effects are most pronounced. Due the metallic behavior,\cite{Jiang_081408_2018} non-significant red-shift in the optical band gap is seen, which suggests an exciton binding energy closer to \SI{0}{\electronvolt} for the ground state exciton. These effects are consistent across all three materials, as inferred from Fig.~\ref{fig:optical_absorption}. However, their impact is slightly less prominent in silverene and copperene due to differences in their single particle levels. The small anisotropy of the optical absorption in enhanced with the quasi-particle excitonic effects showing a higher absorption coefficient for $y$ incident light polarization.

A notable trend in absorption intensity is observed across the materials. Goldene exhibits the highest absorption intensity (see Fig.~\ref{fig:optical_absorption}(a)), particularly around the peak at \SI{2.0}{\electronvolt}, followed by copperene and silverene. This trend can be attributed to the different degree of orbital hybridization and the availability of electronic transitions in each material. Silverene contrasting with cooperene and goldene, shows a lower absorption coefficient in the visible region, which can be justified by it's lower population of valence states around the Fermi level, when compared with Au and Cu counterparts. 

For instance, the peak near \SI{2}{\electronvolt} in goldene is associated with transitions from $d$-orbital-dominated states below the Fermi level to $s$- and $p$-dominated states above the Fermi level. Similar transitions explain the peaks in silverene and copperene, with slight shifts due to differences in bonding strength and electronic configuration.  Moreover, the selective contribution of orbital states --- where $d$-orbitals dominate states below the Fermi level and $s$- and $p$-orbitals dominate states above it --- is critical in shaping their optical response.

\section{Conclusions}

This study employs first-principles calculations to systematically investigate the structural, mechanical, electronic, and optical properties of goldene, silverene, and copperene. Notably, silverene and copperene are introduced here for the first time. Our results confirm that these rectangular 2D materials are dynamically and thermally stable, as evidenced by phonon dispersion and \textit{ab initio} molecular dynamics simulations. The computed cohesion energies indicate that copperene exhibits the highest energetic stability, followed by goldene and silverene, reflecting the influence of distinct bonding environments and reduced dimensionality effects.

Electronic structure analyses reveal that all three materials display metallic band structures, with subtle variations in band dispersion and density of states. Goldene exhibits a sloped band along the X-E path, while silverene and copperene present flatter bands in the same region, suggesting localized electronic states. The electronic transitions and optical behavior are predominantly governed by $d$-orbitals below the Fermi level and $s$- and $p$-orbitals above it.

The mechanical properties demonstrate a nearly isotropic response, with Young's modulus and Poisson's ratio following goldene > copperene > silverene and goldene > silverene > copperene, respectively. Goldene's superior stiffness and ductility are primarily attributed to relativistic effects that strengthen bonding interactions. In contrast, silverene's lower stiffness and copperene's intermediate values highlight their unique mechanical characteristics. 

Despite their metallic nature, goldene, silverene, and copperene exhibit optical responses reminiscent of semiconductors, featuring well-defined absorption onsets and excitonic-like effects that enhance light absorption. It's also presents an optical anisotropy when quasi-particle effects are taken in account, showing a intense optical response for $y$ incident light polarization. These findings underscore the potential of these materials for applications requiring tailored electronic and optical functionalities.

\begin{acknowledgement}
The authors express their gratitude to the National Laboratory for Scientific Computing for providing resources through the Santos Dumont supercomputer and to the ``Centro Nacional de Processamento de Alto Desempenho em S\~ao Paulo'' (CENAPAD-SP, UNICAMP/FINEP - MCTI project) for support related to project 897. Additional resources were provided by Lobo Carneiro HPC (NACAD) at the Federal University of Rio de Janeiro (UFRJ) for project 133. M.L.P.J. acknowledges financial support from the Research Support Foundation of the Federal District (FAP-DF, grant 00193-00001807/2023-16) and the National Council for Scientific and Technological Development (CNPq, grant 444921/2024-9). A.C.D acknowledges financial support from FAP-DF grants 00193-00001817/2023-43 and 00193-00002073/2023-84, CNPq grants 305174/2023-1, 444069/2024-0 and 444431/2024-1. L.A.R.J acknowledges financial support from FAP-DF grant 0193.000942/2015, CNPq grants 307345/ 2021-1 and 350176/2022-1, and FAPDF-PRONEM grant 00193.00001247 /2021-20. A.C.D. and L.A.R.J also acknowledge funding from PDPG-FAPDF-CAPES Centro-Oeste grant number 00193-00000867/2024-94. D. S. G. acknowledges the Center for Computing in Engineering and Sciences at Unicamp for financial support through the FAPESP/CEPID Grant \#2013/08293-7. This paper was published under the CC BY Open Access license through the ACS-CAPES agreement. The authors gratefully acknowledge CAPES for the financial support provided for the publication of this work under the ACS-CAPES agreement.
\end{acknowledgement}

\bibliography{references}

\providecommand{\latin}[1]{#1}
\makeatletter
\providecommand{\doi}
  {\begingroup\let\do\@makeother\dospecials
  \catcode`\{=1 \catcode`\}=2 \doi@aux}
\providecommand{\doi@aux}[1]{\endgroup\texttt{#1}}
\makeatother
\providecommand*\mcitethebibliography{\thebibliography}
\csname @ifundefined\endcsname{endmcitethebibliography}
  {\let\endmcitethebibliography\endthebibliography}{}
\begin{mcitethebibliography}{43}
\providecommand*\natexlab[1]{#1}
\providecommand*\mciteSetBstSublistMode[1]{}
\providecommand*\mciteSetBstMaxWidthForm[2]{}
\providecommand*\mciteBstWouldAddEndPuncttrue
  {\def\EndOfBibitem{\unskip.}}
\providecommand*\mciteBstWouldAddEndPunctfalse
  {\let\EndOfBibitem\relax}
\providecommand*\mciteSetBstMidEndSepPunct[3]{}
\providecommand*\mciteSetBstSublistLabelBeginEnd[3]{}
\providecommand*\EndOfBibitem{}
\mciteSetBstSublistMode{f}
\mciteSetBstMaxWidthForm{subitem}{(\alph{mcitesubitemcount})}
\mciteSetBstSublistLabelBeginEnd
  {\mcitemaxwidthsubitemform\space}
  {\relax}
  {\relax}

\bibitem[Lemme \latin{et~al.}(2022)Lemme, Akinwande, Huyghebaert, and
  Stampfer]{lemme20222d}
Lemme,~M.~C.; Akinwande,~D.; Huyghebaert,~C.; Stampfer,~C. 2D materials for
  future heterogeneous electronics. \emph{Nature communications} \textbf{2022},
  \emph{13}, 1392\relax
\mciteBstWouldAddEndPuncttrue
\mciteSetBstMidEndSepPunct{\mcitedefaultmidpunct}
{\mcitedefaultendpunct}{\mcitedefaultseppunct}\relax
\EndOfBibitem
\bibitem[Xia \latin{et~al.}(2014)Xia, Wang, Xiao, Dubey, and
  Ramasubramaniam]{xia2014two}
Xia,~F.; Wang,~H.; Xiao,~D.; Dubey,~M.; Ramasubramaniam,~A. Two-dimensional
  material nanophotonics. \emph{Nature photonics} \textbf{2014}, \emph{8},
  899--907\relax
\mciteBstWouldAddEndPuncttrue
\mciteSetBstMidEndSepPunct{\mcitedefaultmidpunct}
{\mcitedefaultendpunct}{\mcitedefaultseppunct}\relax
\EndOfBibitem
\bibitem[Deng \latin{et~al.}(2016)Deng, Novoselov, Fu, Zheng, Tian, and
  Bao]{deng2016catalysis}
Deng,~D.; Novoselov,~K.; Fu,~Q.; Zheng,~N.; Tian,~Z.; Bao,~X. Catalysis with
  two-dimensional materials and their heterostructures. \emph{Nature
  nanotechnology} \textbf{2016}, \emph{11}, 218--230\relax
\mciteBstWouldAddEndPuncttrue
\mciteSetBstMidEndSepPunct{\mcitedefaultmidpunct}
{\mcitedefaultendpunct}{\mcitedefaultseppunct}\relax
\EndOfBibitem
\bibitem[Pomerantseva and Gogotsi(2017)Pomerantseva, and
  Gogotsi]{pomerantseva2017two}
Pomerantseva,~E.; Gogotsi,~Y. Two-dimensional heterostructures for energy
  storage. \emph{Nature Energy} \textbf{2017}, \emph{2}, 1--6\relax
\mciteBstWouldAddEndPuncttrue
\mciteSetBstMidEndSepPunct{\mcitedefaultmidpunct}
{\mcitedefaultendpunct}{\mcitedefaultseppunct}\relax
\EndOfBibitem
\bibitem[Geim(2009)]{geim2009graphene}
Geim,~A.~K. Graphene: status and prospects. \emph{science} \textbf{2009},
  \emph{324}, 1530--1534\relax
\mciteBstWouldAddEndPuncttrue
\mciteSetBstMidEndSepPunct{\mcitedefaultmidpunct}
{\mcitedefaultendpunct}{\mcitedefaultseppunct}\relax
\EndOfBibitem
\bibitem[Ranjan \latin{et~al.}(2020)Ranjan, Lee, Kumar, and
  Vinu]{ranjan2020borophene}
Ranjan,~P.; Lee,~J.~M.; Kumar,~P.; Vinu,~A. Borophene: New sensation in
  flatland. \emph{Advanced Materials} \textbf{2020}, \emph{32}, 2000531\relax
\mciteBstWouldAddEndPuncttrue
\mciteSetBstMidEndSepPunct{\mcitedefaultmidpunct}
{\mcitedefaultendpunct}{\mcitedefaultseppunct}\relax
\EndOfBibitem
\bibitem[Carvalho \latin{et~al.}(2016)Carvalho, Wang, Zhu, Rodin, Su, and
  Castro~Neto]{carvalho2016phosphorene}
Carvalho,~A.; Wang,~M.; Zhu,~X.; Rodin,~A.~S.; Su,~H.; Castro~Neto,~A.~H.
  Phosphorene: from theory to applications. \emph{Nature Reviews Materials}
  \textbf{2016}, \emph{1}, 1--16\relax
\mciteBstWouldAddEndPuncttrue
\mciteSetBstMidEndSepPunct{\mcitedefaultmidpunct}
{\mcitedefaultendpunct}{\mcitedefaultseppunct}\relax
\EndOfBibitem
\bibitem[Kara \latin{et~al.}(2012)Kara, Enriquez, Seitsonen, Voon, Vizzini,
  Aufray, and Oughaddou]{kara2012review}
Kara,~A.; Enriquez,~H.; Seitsonen,~A.~P.; Voon,~L. L.~Y.; Vizzini,~S.;
  Aufray,~B.; Oughaddou,~H. A review on silicene—new candidate for
  electronics. \emph{Surface science reports} \textbf{2012}, \emph{67},
  1--18\relax
\mciteBstWouldAddEndPuncttrue
\mciteSetBstMidEndSepPunct{\mcitedefaultmidpunct}
{\mcitedefaultendpunct}{\mcitedefaultseppunct}\relax
\EndOfBibitem
\bibitem[Kashiwaya \latin{et~al.}(2024)Kashiwaya, Shi, Lu, Sangiovanni,
  Greczynski, Magnuson, Andersson, Rosen, and Hultman]{kashiwaya2024synthesis}
Kashiwaya,~S.; Shi,~Y.; Lu,~J.; Sangiovanni,~D.~G.; Greczynski,~G.;
  Magnuson,~M.; Andersson,~M.; Rosen,~J.; Hultman,~L. Synthesis of goldene
  comprising single-atom layer gold. \emph{Nature Synthesis} \textbf{2024},
  1--8\relax
\mciteBstWouldAddEndPuncttrue
\mciteSetBstMidEndSepPunct{\mcitedefaultmidpunct}
{\mcitedefaultendpunct}{\mcitedefaultseppunct}\relax
\EndOfBibitem
\bibitem[Sharma \latin{et~al.}(2022)Sharma, Pasricha, Weston, Blanton, and
  Jagannathan]{sharma2022synthesis}
Sharma,~S.~K.; Pasricha,~R.; Weston,~J.; Blanton,~T.; Jagannathan,~R. Synthesis
  of self-assembled single atomic layer gold crystals-goldene. \emph{ACS
  applied materials \& interfaces} \textbf{2022}, \emph{14}, 54992--55003\relax
\mciteBstWouldAddEndPuncttrue
\mciteSetBstMidEndSepPunct{\mcitedefaultmidpunct}
{\mcitedefaultendpunct}{\mcitedefaultseppunct}\relax
\EndOfBibitem
\bibitem[Mortazavi(2024)]{mortazavi2024goldene}
Mortazavi,~B. Goldene: An Anisotropic Metallic Monolayer with Remarkable
  Stability and Rigidity and Low Lattice Thermal Conductivity. \emph{Materials}
  \textbf{2024}, \emph{17}, 2653\relax
\mciteBstWouldAddEndPuncttrue
\mciteSetBstMidEndSepPunct{\mcitedefaultmidpunct}
{\mcitedefaultendpunct}{\mcitedefaultseppunct}\relax
\EndOfBibitem
\bibitem[Zhao \latin{et~al.}(2024)Zhao, Zhang, Zhu, Jiang, and
  Zheng]{zhao2024electrical}
Zhao,~S.; Zhang,~H.; Zhu,~M.; Jiang,~L.; Zheng,~Y. Electrical conductivity of
  goldene. \emph{Physical Review B} \textbf{2024}, \emph{110}, 085111\relax
\mciteBstWouldAddEndPuncttrue
\mciteSetBstMidEndSepPunct{\mcitedefaultmidpunct}
{\mcitedefaultendpunct}{\mcitedefaultseppunct}\relax
\EndOfBibitem
\bibitem[Nguyen \latin{et~al.}(2025)Nguyen, Nguyen, Hieu, Phuc, and
  Nguyen]{nguyen2025goldene}
Nguyen,~S.; Nguyen,~C.; Hieu,~N.; Phuc,~H.; Nguyen,~C. Goldene: A promising
  electrode for achieving ultra-low Schottky contact in metal--semiconductor
  Goldene/MX2 (M= Mo, W; X= S, Se) heterostructure. \emph{Materials Science in
  Semiconductor Processing} \textbf{2025}, \emph{185}, 108986\relax
\mciteBstWouldAddEndPuncttrue
\mciteSetBstMidEndSepPunct{\mcitedefaultmidpunct}
{\mcitedefaultendpunct}{\mcitedefaultseppunct}\relax
\EndOfBibitem
\bibitem[Abidi and Koskinen(2024)Abidi, and Koskinen]{abidi2024gentle}
Abidi,~K.~R.; Koskinen,~P. Gentle tension stabilizes atomically thin
  metallenes. \emph{Nanoscale} \textbf{2024}, \emph{16}, 19649--19655\relax
\mciteBstWouldAddEndPuncttrue
\mciteSetBstMidEndSepPunct{\mcitedefaultmidpunct}
{\mcitedefaultendpunct}{\mcitedefaultseppunct}\relax
\EndOfBibitem
\bibitem[J{\'u}nior \latin{et~al.}(2025)J{\'u}nior, dos Santos, Junior, and
  Galvao]{junior2025does}
J{\'u}nior,~M. L.~P.; dos Santos,~E. J.~A.; Junior,~L. A.~R.; Galvao,~D. How
  does Goldene Stack? \emph{Materials Horizons} \textbf{2025}, \relax
\mciteBstWouldAddEndPunctfalse
\mciteSetBstMidEndSepPunct{\mcitedefaultmidpunct}
{}{\mcitedefaultseppunct}\relax
\EndOfBibitem
\bibitem[Berdiyorov and Aissa(2025)Berdiyorov, and Aissa]{berdiyorov2025robust}
Berdiyorov,~G.; Aissa,~B. Robust Conductivity of Goldene Against Structural
  Defects and Mechanical Deformations: A First-Principles Study. \textbf{2025},
  \relax
\mciteBstWouldAddEndPunctfalse
\mciteSetBstMidEndSepPunct{\mcitedefaultmidpunct}
{}{\mcitedefaultseppunct}\relax
\EndOfBibitem
\bibitem[ABIDI and Koskinen(2025)ABIDI, and Koskinen]{Abidi_1_2025}
ABIDI,~K.~R.; Koskinen,~P. Electronic and structural properties of atomically
  thin metallenes. \emph{Electronic Structure} \textbf{2025}, \relax
\mciteBstWouldAddEndPunctfalse
\mciteSetBstMidEndSepPunct{\mcitedefaultmidpunct}
{}{\mcitedefaultseppunct}\relax
\EndOfBibitem
\bibitem[Ramachandran \latin{et~al.}(2024)Ramachandran, Ali, Butt, Zheng,
  Deader, and Rezeq]{ramachandran2024gold}
Ramachandran,~T.; Ali,~A.; Butt,~H.; Zheng,~L.; Deader,~F.~A.; Rezeq,~M. Gold
  on the horizon: unveiling the chemistry, applications and future prospects of
  2D monolayers of gold nanoparticles (Au-NPs). \emph{Nanoscale Advances}
  \textbf{2024}, \emph{6}, 5478--5510\relax
\mciteBstWouldAddEndPuncttrue
\mciteSetBstMidEndSepPunct{\mcitedefaultmidpunct}
{\mcitedefaultendpunct}{\mcitedefaultseppunct}\relax
\EndOfBibitem
\bibitem[Ono and Yoshioka(2024)Ono, and Yoshioka]{ono2024breakdown}
Ono,~S.; Yoshioka,~H. Breakdown of continuum elasticity due to electronic
  effects in gold nanotubes. \emph{arXiv preprint arXiv:2411.08289}
  \textbf{2024}, \relax
\mciteBstWouldAddEndPunctfalse
\mciteSetBstMidEndSepPunct{\mcitedefaultmidpunct}
{}{\mcitedefaultseppunct}\relax
\EndOfBibitem
\bibitem[Kresse and Hafner(1993)Kresse, and Hafner]{kresse1993ab}
Kresse,~G.; Hafner,~J. Ab initio molecular dynamics for open-shell transition
  metals. \emph{Physical Review B} \textbf{1993}, \emph{48}, 13115\relax
\mciteBstWouldAddEndPuncttrue
\mciteSetBstMidEndSepPunct{\mcitedefaultmidpunct}
{\mcitedefaultendpunct}{\mcitedefaultseppunct}\relax
\EndOfBibitem
\bibitem[Kresse and Furthm{\"u}ller(1996)Kresse, and
  Furthm{\"u}ller]{kresse1996efficient}
Kresse,~G.; Furthm{\"u}ller,~J. Efficient iterative schemes for ab initio
  total-energy calculations using a plane-wave basis set. \emph{Physical review
  B} \textbf{1996}, \emph{54}, 11169\relax
\mciteBstWouldAddEndPuncttrue
\mciteSetBstMidEndSepPunct{\mcitedefaultmidpunct}
{\mcitedefaultendpunct}{\mcitedefaultseppunct}\relax
\EndOfBibitem
\bibitem[Perdew \latin{et~al.}(1996)Perdew, Burke, and
  Ernzerhof]{perdew1996generalized}
Perdew,~J.~P.; Burke,~K.; Ernzerhof,~M. Generalized gradient approximation made
  simple. \emph{Physical review letters} \textbf{1996}, \emph{77}, 3865\relax
\mciteBstWouldAddEndPuncttrue
\mciteSetBstMidEndSepPunct{\mcitedefaultmidpunct}
{\mcitedefaultendpunct}{\mcitedefaultseppunct}\relax
\EndOfBibitem
\bibitem[Bl{\"o}chl(1994)]{Blochl_17953_1994}
Bl{\"o}chl,~P.~E. Projector Augmented-Wave Method. \emph{Phys. Rev. B}
  \textbf{1994}, \emph{50}, 17953--17979\relax
\mciteBstWouldAddEndPuncttrue
\mciteSetBstMidEndSepPunct{\mcitedefaultmidpunct}
{\mcitedefaultendpunct}{\mcitedefaultseppunct}\relax
\EndOfBibitem
\bibitem[Kresse and Joubert(1999)Kresse, and Joubert]{Kresse_1758_1999}
Kresse,~G.; Joubert,~D. From Ultrasoft Pseudopotentials to the Projector
  Augmented-Wave Method. \emph{Phys. Rev. B} \textbf{1999}, \emph{59},
  1758--1775\relax
\mciteBstWouldAddEndPuncttrue
\mciteSetBstMidEndSepPunct{\mcitedefaultmidpunct}
{\mcitedefaultendpunct}{\mcitedefaultseppunct}\relax
\EndOfBibitem
\bibitem[Monkhorst and Pack(1976)Monkhorst, and Pack]{monkhorst1976special}
Monkhorst,~H.~J.; Pack,~J.~D. Special points for Brillouin-zone integrations.
  \emph{Physical review B} \textbf{1976}, \emph{13}, 5188\relax
\mciteBstWouldAddEndPuncttrue
\mciteSetBstMidEndSepPunct{\mcitedefaultmidpunct}
{\mcitedefaultendpunct}{\mcitedefaultseppunct}\relax
\EndOfBibitem
\bibitem[Togo and Tanaka(2015)Togo, and Tanaka]{Togo_1_2015}
Togo,~A.; Tanaka,~I. First principles phonon calculations in materials science.
  \emph{Scripta Materialia} \textbf{2015}, \emph{108}, 1--5\relax
\mciteBstWouldAddEndPuncttrue
\mciteSetBstMidEndSepPunct{\mcitedefaultmidpunct}
{\mcitedefaultendpunct}{\mcitedefaultseppunct}\relax
\EndOfBibitem
\bibitem[Gajdoš \latin{et~al.}(2006)Gajdoš, Hummer, Kresse, Furthm\"{u}ller,
  and Bechstedt]{Gajdos_045112_2006}
Gajdoš,~M.; Hummer,~K.; Kresse,~G.; Furthm\"{u}ller,~J.; Bechstedt,~F. Linear
  optical properties in the projector-augmented wave methodology.
  \emph{Physical Review B} \textbf{2006}, \emph{73}\relax
\mciteBstWouldAddEndPuncttrue
\mciteSetBstMidEndSepPunct{\mcitedefaultmidpunct}
{\mcitedefaultendpunct}{\mcitedefaultseppunct}\relax
\EndOfBibitem
\bibitem[Hoover \latin{et~al.}(1982)Hoover, Ladd, and Moran]{Hoover_1818_1982}
Hoover,~W.~G.; Ladd,~A. J.~C.; Moran,~B. High-Strain-Rate Plastic Flow Studied
  via Nonequilibrium Molecular Dynamics. \emph{Physical Review Letters}
  \textbf{1982}, \emph{48}, 1818–1820\relax
\mciteBstWouldAddEndPuncttrue
\mciteSetBstMidEndSepPunct{\mcitedefaultmidpunct}
{\mcitedefaultendpunct}{\mcitedefaultseppunct}\relax
\EndOfBibitem
\bibitem[Evans(1983)]{Evans_3297_1983}
Evans,~D.~J. Computer ‘“experiment”’ for nonlinear thermodynamics of
  Couette flow. \emph{The Journal of Chemical Physics} \textbf{1983},
  \emph{78}, 3297–3302\relax
\mciteBstWouldAddEndPuncttrue
\mciteSetBstMidEndSepPunct{\mcitedefaultmidpunct}
{\mcitedefaultendpunct}{\mcitedefaultseppunct}\relax
\EndOfBibitem
\bibitem[Salpeter and Bethe(1951)Salpeter, and Bethe]{Salpeter_1232_1951}
Salpeter,~E.~E.; Bethe,~H.~A. A Relativistic Equation for Bound-State Problems.
  \emph{Phys. Rev.} \textbf{1951}, \emph{84}, 1232--1242\relax
\mciteBstWouldAddEndPuncttrue
\mciteSetBstMidEndSepPunct{\mcitedefaultmidpunct}
{\mcitedefaultendpunct}{\mcitedefaultseppunct}\relax
\EndOfBibitem
\bibitem[Dias \latin{et~al.}(2023)Dias, Silveira, and Qu]{dias2023wantibexos}
Dias,~A.~C.; Silveira,~J.~F.; Qu,~F. WanTiBEXOS: A Wannier based Tight Binding
  code for electronic band structure, excitonic and optoelectronic properties
  of solids. \emph{Computer Physics Communications} \textbf{2023}, \emph{285},
  108636\relax
\mciteBstWouldAddEndPuncttrue
\mciteSetBstMidEndSepPunct{\mcitedefaultmidpunct}
{\mcitedefaultendpunct}{\mcitedefaultseppunct}\relax
\EndOfBibitem
\bibitem[Heyd and Scuseria(2004)Heyd, and Scuseria]{heyd_1187_2004}
Heyd,~J.; Scuseria,~G.~E. Efficient hybrid density functional calculations in
  solids: Assessment of the Heyd--Scuseria--Ernzerhof screened Coulomb hybrid
  functional. \emph{The Journal of chemical physics} \textbf{2004}, \emph{121},
  1187--1192\relax
\mciteBstWouldAddEndPuncttrue
\mciteSetBstMidEndSepPunct{\mcitedefaultmidpunct}
{\mcitedefaultendpunct}{\mcitedefaultseppunct}\relax
\EndOfBibitem
\bibitem[Mostofi \latin{et~al.}(2008)Mostofi, Yates, Lee, Souza, Vanderbilt,
  and Marzari]{mostofi2008wannier90}
Mostofi,~A.~A.; Yates,~J.~R.; Lee,~Y.-S.; Souza,~I.; Vanderbilt,~D.;
  Marzari,~N. wannier90: A tool for obtaining maximally-localised Wannier
  functions. \emph{Computer physics communications} \textbf{2008}, \emph{178},
  685--699\relax
\mciteBstWouldAddEndPuncttrue
\mciteSetBstMidEndSepPunct{\mcitedefaultmidpunct}
{\mcitedefaultendpunct}{\mcitedefaultseppunct}\relax
\EndOfBibitem
\bibitem[Rozzi \latin{et~al.}(2006)Rozzi, Varsano, Marini, Gross, and
  Rubio]{Rozzi_205119_2006}
Rozzi,~C.~A.; Varsano,~D.; Marini,~A.; Gross,~E. K.~U.; Rubio,~A. Exact Coulomb
  cutoff technique for supercell calculations. \emph{Physical Review B}
  \textbf{2006}, \emph{73}\relax
\mciteBstWouldAddEndPuncttrue
\mciteSetBstMidEndSepPunct{\mcitedefaultmidpunct}
{\mcitedefaultendpunct}{\mcitedefaultseppunct}\relax
\EndOfBibitem
\bibitem[Pyykko and Desclaux(1979)Pyykko, and Desclaux]{pyykko1979relativity}
Pyykko,~P.; Desclaux,~J.~P. Relativity and the periodic system of elements.
  \emph{Accounts of Chemical Research} \textbf{1979}, \emph{12}, 276--281\relax
\mciteBstWouldAddEndPuncttrue
\mciteSetBstMidEndSepPunct{\mcitedefaultmidpunct}
{\mcitedefaultendpunct}{\mcitedefaultseppunct}\relax
\EndOfBibitem
\bibitem[Schwerdtfeger(2002)]{schwerdtfeger2002relativistic}
Schwerdtfeger,~P. Relativistic effects in properties of gold. \emph{Heteroatom
  Chemistry: An International Journal of Main Group Elements} \textbf{2002},
  \emph{13}, 578--584\relax
\mciteBstWouldAddEndPuncttrue
\mciteSetBstMidEndSepPunct{\mcitedefaultmidpunct}
{\mcitedefaultendpunct}{\mcitedefaultseppunct}\relax
\EndOfBibitem
\bibitem[Mounet and Marzari(2005)Mounet, and Marzari]{mounet2005first}
Mounet,~N.; Marzari,~N. First-principles determination of the structural,
  vibrational and thermodynamic properties of diamond, graphite, and
  derivatives. \emph{Physical Review B—Condensed Matter and Materials
  Physics} \textbf{2005}, \emph{71}, 205214\relax
\mciteBstWouldAddEndPuncttrue
\mciteSetBstMidEndSepPunct{\mcitedefaultmidpunct}
{\mcitedefaultendpunct}{\mcitedefaultseppunct}\relax
\EndOfBibitem
\bibitem[Scholes and Rumbles(2006)Scholes, and Rumbles]{scholes2006excitons}
Scholes,~G.~D.; Rumbles,~G. Excitons in nanoscale systems. \emph{Nature
  materials} \textbf{2006}, \emph{5}, 683--696\relax
\mciteBstWouldAddEndPuncttrue
\mciteSetBstMidEndSepPunct{\mcitedefaultmidpunct}
{\mcitedefaultendpunct}{\mcitedefaultseppunct}\relax
\EndOfBibitem
\bibitem[Wang \latin{et~al.}(2019)Wang, Chen, Yuan, Rong, Feng, Muzammil, Yu,
  Zhang, and Zhan]{wang2019dhq}
Wang,~X.; Chen,~L.; Yuan,~Z.; Rong,~J.; Feng,~J.; Muzammil,~I.; Yu,~X.;
  Zhang,~Y.; Zhan,~Z. DHQ-graphene: a novel two-dimensional defective graphene
  for corrosion-resistant coating. \emph{Journal of Materials Chemistry A}
  \textbf{2019}, \emph{7}, 8967--8974\relax
\mciteBstWouldAddEndPuncttrue
\mciteSetBstMidEndSepPunct{\mcitedefaultmidpunct}
{\mcitedefaultendpunct}{\mcitedefaultseppunct}\relax
\EndOfBibitem
\bibitem[Dias \latin{et~al.}(2021)Dias, Bragan\c{c}a, de~Mendon\c{c}a, and
  Da~Silva]{Dias_3265_2021}
Dias,~A.~C.; Bragan\c{c}a,~H.; de~Mendon\c{c}a,~J. P.~A.; Da~Silva,~J. L.~F.
  Excitonic Effects on Two-Dimensional Transition-Metal Dichalcogenide
  Monolayers: Impact on Solar Cell Efficiency. \emph{ACS Applied Energy
  Materials} \textbf{2021}, \emph{4}, 3265–3278\relax
\mciteBstWouldAddEndPuncttrue
\mciteSetBstMidEndSepPunct{\mcitedefaultmidpunct}
{\mcitedefaultendpunct}{\mcitedefaultseppunct}\relax
\EndOfBibitem
\bibitem[Cavalheiro~Dias \latin{et~al.}(2024)Cavalheiro~Dias,
  Almeida~Cornélio, Piotrowski, Ribeiro~Júnior, de~Oliveira~Bastos,
  Caldeira~R\^ego, and Guedes-Sobrinho]{Dias_8572_2024}
Cavalheiro~Dias,~A.; Almeida~Cornélio,~C.~D.; Piotrowski,~M.~J.;
  Ribeiro~Júnior,~L.~A.; de~Oliveira~Bastos,~C.~M.; Caldeira~R\^ego,~C.~R.;
  Guedes-Sobrinho,~D. Can 2D Carbon Allotropes Be Used as Photovoltaic
  Absorbers in Solar Harvesting Devices? \emph{ACS Applied Energy Materials}
  \textbf{2024}, \emph{7}, 8572–8582\relax
\mciteBstWouldAddEndPuncttrue
\mciteSetBstMidEndSepPunct{\mcitedefaultmidpunct}
{\mcitedefaultendpunct}{\mcitedefaultseppunct}\relax
\EndOfBibitem
\bibitem[Jiang \latin{et~al.}(2018)Jiang, Li, Zhang, and
  Duan]{Jiang_081408_2018}
Jiang,~Z.; Li,~Y.; Zhang,~S.; Duan,~W. Realizing an intrinsic excitonic
  insulator by decoupling exciton binding energy from the minimum band gap.
  \emph{Physical Review B} \textbf{2018}, \emph{98}\relax
\mciteBstWouldAddEndPuncttrue
\mciteSetBstMidEndSepPunct{\mcitedefaultmidpunct}
{\mcitedefaultendpunct}{\mcitedefaultseppunct}\relax
\EndOfBibitem
\end{mcitethebibliography}

\end{document}